\documentclass{emulateapj}
\usepackage{epstopdf}
\newcommand{\msun}{$\rm M_{\sun}$}
\newcommand{\rsun}{$\rm R_{\sun}$}
\newcommand{\rearth}{${\rm R_{\earth}} $}
\newcommand{\mearth}{$\rm M_{\earth}$}
\newcommand{\hho}{H$_2$O}

\begin{document}

\shorttitle{The GJ1214 Super-Earth System}
\shortauthors{Berta et al.}

\title{The GJ1214 Super-Earth System: Stellar Variability, New Transits, and a Search for Additional Planets}

\author{Zachory~K.~Berta, David~Charbonneau, Jacob~Bean\altaffilmark{1},  Jonathan~Irwin, Christopher~J.~Burke and Jean-Michel~D\'esert}
\affil{Harvard-Smithsonian Center for Astrophysics, 60 Garden St.,
  Cambridge, MA 02138, USA}
\altaffiltext{1}{Sagan Fellow}
\email{zberta@cfa.harvard.edu}

\author{Philip~Nutzman}
\affil{Department of Astronomy and Astrophysics, University of California, Santa Cruz, CA 95064, USA}

\author{Emilio~E.~Falco}
\affil{Fred Lawrence Whipple Observatory, Smithsonian Astrophysical Observatory, 670 Mount Hopkins Road,
  Amado, AZ 85645, USA}

\begin{abstract}
The super-Earth GJ1214b transits a nearby M dwarf that exhibits $1\%$ intrinsic variability in the near-infrared. Here, we analyze new observations to refine the physical properties of both the star and planet. We present three years of out-of-transit photometric monitoring of the stellar host GJ1214 from the MEarth Observatory and find the rotation period to be long, mostly likely an integer multiple of 53 days, suggesting low levels of magnetic activity and an old age for the system. We show such variability will not pose significant problems to ongoing studies of the planet's atmosphere with transmission spectroscopy. We analyze 2 high-precision transit light curves from ESO's Very Large Telescope along with 7 others from the MEarth and FLWO 1.2 meter telescopes, finding physical parameters for the planet that are consistent with previous work. The VLT light curves show tentative evidence for spot occultations during transit. Using two years of MEarth light curves, we place limits on additional transiting planets around GJ1214 with periods out to the habitable zone of the system. We also improve upon the previous photographic $V$-band estimate for the star, finding $V=14.71\pm 0.03$. 

\end{abstract}

\section{Introduction}
\label{introduction}

The transiting exoplanet GJ1214b offers an unparalleled opportunity to explore the physical properties of super-Earth planets. With a mass ($M_p$ = 6.6 \mearth) and radius ($R_p$ = 2.7 \rearth) between those of Earth and Neptune, and a likely equilibrium temperature ($T_{eq} = 500K$) cooler than for most transiting planets, GJ1214b represents an intriguing new kind of world with no Solar System analog \citep{charbonneau.2009.stnls}. Given intrinsic degeneracies in the mass-radius diagram in this regime \citep{seager.2007.mrse, adams.2008.optamrsewma,rogers.2010.fqdeic}, the bulk composition of the planet cannot be uniquely determined from current measurements of the mass and radius alone. For example, \citet{rogers.2010.tpol1} can explain the observed mass and radius to within $1\sigma$ with any of three generic physical models: (i)  a mini-Neptune that accreted and maintained a low-mass H/He layer from the primordial nebula, (ii) a superfluid water-world with a sublimating ${\rm H_2O}$ envelope, or (iii) a rocky planet with an H-dominated atmosphere formed by recent outgassing. { Detailed calculations of GJ1214b's thermal evolution by \citet{2010arXiv1010.0277N} favor a metal-enriched H/He/\hho~envelope, finding that a water-only atmosphere would require an implausibly large water-to-rock ratio in the planet's interior.}

Fortunately, because GJ1214b transits a very nearby (13 pc), bright ($K=8.8$), low-mass M dwarf (0.16 \msun), it is amenable to follow-up observations that could distinguish among these hypotheses. In particular, the large ($D=1.4\%$) transit depth enables transit studies of the planet's atmosphere. \citet{miller-ricci.2009.assdbhha} show that measuring the amplitude of the planet's transmission spectrum (i.e., the wavelength-dependence of the transit depth $\Delta D(\lambda)$ caused by absorption at the limb of the planet) constrains the mean molecular weight of its atmosphere and, in turn, the hydrogen content of its outer envelope. Cases (i) or (iii) of \citet{rogers.2010.tpol1} would produce $\Delta D(\lambda)\approx0.1\%$ variations in the transit depth across wavelengths accessible from the ground as well as {\em Hubble} and {\em Spitzer Space Telescopes}, while case (ii)'s dense atmosphere would result in variations below the sensitivity of current instruments  \citep{miller-ricci.2010.nats1}.

Providing a potential complication, however, the host star GJ1214 shows roughly sinusoidal photometric modulations that are presumably due to an asymmetric distribution of spots on a rotating star. Such spots can bias planetary parameters as measured from transit light curves whether or not they are occulted by the planet \citep[e.g.][]{pont.2007.hsttppt1mrs, desert.2011.tse1isom}, partially decoupling the observed transit depth $D$ from the actual planet-to-star radius ratio $R_p/R_{\star}$. Of particular importance for transmission spectroscopy studies, the change in transit depth induced by spots can vary with both time and wavelength, potentially mimicking the signal of a planetary atmosphere.

Stellar spots have been observed in several transiting exoplanet systems around active stars \citep[see][]{strassmeier.2009.s}. {\em Hubble Space Telescope} ({\em HST}) photometry \citep{rabus.2009.csstpts} and later ground-based follow-up \citep{dittmann.2009.tdsdctepfgedtpsat} of TrES-1b has shown evidence for spot occultations in transit light curves. The high photometric precision and continuous coverage provided by the { CoRoT} satellite enabled detailed modeling of spotty transit and out-of-transit light curves for the hot Jupiters { CoRoT-2b} \citep{wolter.2009.tmscpsswpt, czesla.2009.saaseep, huber.2010.pemcedrsm} and { CoRoT-4b} \citep{aigrain.2009.npcdpp,lanza.2009.parpsc}. For the former, joint fits to the transit and out-of-transit flux showed that initial estimates of the planet's $R_p/R_{\star}$ were 3\% (9$\sigma$) too low \citep{czesla.2009.saaseep}. The interpretation of the transiting super-Earth { CoRoT-7b} is obfuscated by the fact that both the transit depth and the reflex motion are well below the amplitude of activity-induced modulations \citep{leger.2009.tefcsmvcfswmr,queloz.2009.cpsos}. Reanalyses of the { CoRoT-7} radial velocities find changing values for the mass of CoRoT-7b \citep{hatzes.2010.iirvvc,ferraz-mello.2010.pmdcsoasccs,lanza.2010.parrvvpsc} and call into question the significance of the mass measurements for both { CoRoT-7b} and the claimed outer planet { CoRoT-7c} \citep{pont.2010.rrepac}.

Like GJ1214b, the well-studied hot Jupiter HD189733b \citep{bouchy.2005.emstjivjtbs1} is an ideal system for characterization studies, but requires corrections for stellar activity. The host HD189733 is an active K2 dwarf \citep{moutou.2007.sotpsd1} with 2\% peak-to-peak variability in the optical \citep{croll.2007.ls1sstmsp,miller-ricci.2008.msptes1ptmtaas}. \citet{henry.2008.rpps1} undertook a long-term photometric monitoring campaign from which they measured the 12 day stellar rotation period of HD189733. Extrapolation from their out-of-eclipse photometric spot characterization was useful for interpreting transmission spectroscopy results of individual transits from {\em Hubble} \citep[][]{pont.2007.hsttppt1mrs, swain.2008.pmaep} and measurements of the thermal phase curve from {\em Spitzer} \citep{knutson.2007.dcep1,knutson.2009.mcdcp1}. Understanding the time-variable surface of the star was even more crucial for broadband transmission spectroscopy studies that rely on comparing transit depths at different epochs \citep[e.g.,][]{desert.2009.scmate1,desert.2010.tsehisom,sing.2009.tse1iswfhwn}; interpretation of these data rely heavily on the photometric monitoring of \citet{henry.2008.rpps1}.

To aid ongoing and future studies of GJ1214b, we present new data (\S\ref{Observations}) to characterize the star GJ1214's variability and estimate its rotation period (\S\ref{Rotation}). We compare the measured variability to a simultaneous analysis of 2 high-precision transit light curves from ESO's Very Large Telescope with 7 other new or previously published transits (\S\ref{Transits}). Additionally, we place upper limits on the radii of other possible transiting planets in the system (\S\ref{Search}) and present a refined estimate of the star's $V$ flux, which bears directly upon its metallicity as estimated using $M_K$ and $V-K$ relations. Finally, we discuss the implications of the measured variability for the properties of the star and for transmission spectroscopy studies of GJ1214b's atmosphere (\S\ref{Discussion}).

We also note the following correction. In \citet{charbonneau.2009.stnls}, we quoted a systemic radial velocity for GJ1214 that had a typo in the sign; the actual velocity is $\gamma=+21.1\pm1.0$ km s$^{-1}$ (i.e. a redshift).

\section{Observations and Data Reduction}
\label{Observations}

\subsection{MEarth Photometry}

We monitored the brightness of the GJ1214 system at a variety of cadences with the MEarth Observatory at Mt. Hopkins, AZ throughout the 2008, 2009, and 2010 spring observing seasons. As described in \citet{nutzman.2008.dcgtshpod}, the MEarth Observatory was designed to detect transiting exoplanets around nearby M dwarfs, and consists of eight identical 40-cm telescopes on German Equatorial mounts in a single enclosure at the Fred Lawrence Whipple Observatory (FLWO).  Each telescope is equipped with a thinned, back-illuminated 2048x2048 CCD with a pixel scale of 0.757''/pixel for a 26' field of view. For the bulk of the data presented in this work telescopes were equipped with a fixed, custom, 715 nm long-pass filter; the response is similar to a combination of the Sloan $i + z$ bandpasses and will be hereafter referred to as the ``MEarth'' bandpass. The MEarth Observatory is almost fully automated and operates on every clear night, observing target stars selected from a list of 2000 nearby late M dwarfs \citep{nutzman.2008.dcgtshpod}. In typical operating mode, each telescope observes its own list of 20-30 stars per night when they are above airmass 2 with the cadence and exposure times necessary to detect a transiting planet as small as 2 \rearth~in each star's habitable zone. 

Light curves are extracted automatically from MEarth images by a modified version of the Monitor pipeline \citep{irwin.2007.mpdplcp}, using nightly flat field (dawn and dusk), dark, and bias exposures for calibration. { A differential photometry correction for each frame was calculated from a robust, weighted fit to 78 automatically selected field comparison stars within 2.3 instrumental magnitudes of GJ1214. The mean MEarth - $K$ color of these stars is 0.98; none were as red as GJ1214 (MEarth - $K$ = 2.19).} Predicted uncertainties for each measurement are calculated from a standard CCD noise model. 

GJ1214 was observed with three main cadences. ``Low'' cadence (20-30 minutes between exposures) was that associated with the normal survey mode, and was employed in the 2008 and 2009 seasons. ``Medium'' cadence (5-10 minutes) was implemented after the discovery of GJ1214b and was intended to boost sensitivity both to other transiting planets and to characterizing the out-of-transit variability of the star. ``High'' cadence (40 seconds) was employed at predicted times of transit to determine the system parameters. High cadence transits were observed simultaneously with 7 or 8 MEarth telescopes for greater precision, as the systematic noise sources and scintillation patterns among pairs of MEarth telescopes appear to be largely uncorrelated, so the S/N improvement scales with the square root of the number of telescopes. In Table~\ref{tab-LC}, we present one new MEarth transit light curve, along with the four MEarth and two KeplerCam light curves that were analyzed but not made electronically available in  \citet{charbonneau.2009.stnls}.

While we include data from 2008 for the rotation analysis, we caution that these observations took place during MEarth's early commissioning, before the observing strategy and software were finalized. Changes to the telescope throughout the season may have corrupted the season-long stability. Importantly, a field acquisition loop designed to mitigate flat-fielding errors by bringing each star back to the same pixel was not implemented until the late spring of 2008. During the 2008 season, a Bessell-prescription I filter \citep{bessell.1990.up} was used instead of the custom MEarth bandpass.

\begin{deluxetable*}{lrrlcc}
\tabletypesize{\normalsize}
\tablecaption{\label{tab-LC} Transit Light Curves}
\tablecolumns{6}
\tablehead{\colhead{Mid-exposure Time\tablenotemark{a}} & \colhead{Relative Flux\tablenotemark{b}} & \colhead{Error\tablenotemark{c}} & \colhead{Airmass} & \colhead{Instrument} & \colhead{Cycle\tablenotemark{d}}}
\startdata
2454964.8926699 & 0.99622 & 0.00317 & 1.1197 & MEARTH & 0 \\
2454964.8933879 & 0.99777 & 0.00321 & 1.1198 & MEARTH & 0 \\
2454964.8941049 & 1.00578 & 0.00325 & 1.1199 & MEARTH & 0 \\

\nodata \\
2454980.7148766 & 0.99733 & 0.00185 & 1.6190 & FLWO & 9 \\
2454980.7153966 & 1.00044 & 0.00185 & 1.6140 & FLWO & 9 \\
2454980.7158946 & 0.99959 & 0.00185 & 1.6080 & FLWO & 9 \\

\nodata \\
2455315.7660750 & 0.99786 & 0.00014 & 1.2125 & VLT & 221 \\
2455315.7668492 & 0.99762 & 0.00014 & 1.2105 & VLT & 221 \\
2455315.7702432 & 0.99763 & 0.00012 & 1.2020 & VLT & 221 \\

\enddata
\tablecomments{This table is presented in its entirety in the electronic edition; a portion is shown here for guidance regarding its form and content.}
\tablenotetext{a}{Times are given as ${\rm BJD_{TDB}}$, Barycentric Julian Dates in the Barycentric Dynamical Time system \citep{eastman.2010.abtmahbjd}.}
\tablenotetext{b}{Differential photometry corrections have been applied, but additional systematic corrections (see text) have not. Each light curve has been divided by the median out-of-transit flux. }
\tablenotetext{c}{Theoretical $1\sigma$ errors have been calculated from a standard CCD noise model.}
\tablenotetext{d}{Time measured from the reference epoch in units of the orbital period.}

\end{deluxetable*}

\subsection{V-band Photometry from KeplerCam.}

An independent out-of-transit light curve was obtained through Harris $V$ and $I$ filters\footnote{http://www.sao.arizona.edu/FLWO/48/CCD.filters.html.} with KeplerCam on the 1.2 meter reflector at FLWO atop Mt. Hopkins, AZ. The observations discussed here were gathered in service mode from 26 March 2010 until { 17 June 2010}, after which date the mirror was taken off and put back on the telescope, introducing an uncorrectable systematic offset to the field light curves so later data had to be discarded.

Given the large night-to-night positional shifts of the field, we made an effort to quantify and ameliorate flat-fielding errors by sampling multiple regions of the detector, with each observation consisting of a set of three exposures offset by 3 arcminute dithers. Individual exposures had theoretical noise limits ranging from 0.3\% to 2\%, but the scatter among dither points suggested that calibration errors from flat-fielding introduced a 1\% noise floor to the light curve. Dark sky flats were generated and corrected over time for changes at high spatial frequencies (i.e. dust donuts) by nightly dome flat exposures.

We measured calibrated $V$ and $I$ magnitudes (Table~\ref{tab-vi}) to improve on previously published photographic estimates \citep{lepine.2005.cnswapmlt0lc}. Standard fields \citep{landolt.1992.upssmr1ace} were observed on the nights of 26, 27, and 28 March. Conditions were clear, although seeing as poor as 10" FWHM was witnessed. We estimate the calibration uncertainties for the nights from the scatter in multiple standard exposures. 

\subsection{Light Curves from VLT-FORS2}

Spectra of GJ1214 and 6 comparison stars were gathered during three transits of GJ1214b using ESO Director's Discretionary Time on the VLT (Prog. ID \#284.C-5042 and 285.C-5019).  As described by \citet{bean.2010.temp}, the primary purpose for obtaining these data was to measure the transmission spectrum of GJ1214b's atmosphere by generating multiwavelength transit light curves and determining the wavelength-dependence of the transit depth. In this work, we generate and analyze high precision ``white'' light curves by summing together all the photons collected in each spectrum.

Observations were performed in queue mode with the multiobject, low dispersion spectrograph FORS2 \citep{appenzeller.1998.scffoi} on VLT/UT1. The spectrograph was configured with the 600z+23 grism with a central wavelength of 900 nm and the red-sensitive (MIT) CCD in the standard 2x2 read mode. Exposure times were 20-40 seconds, and the readout time was 37 seconds. A custom slit mask was used; each slit was a rectangle 12'' in the dispersion direction and 15-30'' in the spatial direction, small enough to isolate GJ1214 and the comparison stars but large enough that changing slit losses due to variable seeing were negligible. Wavelength calibration exposures with a He, Ne, Ar emission lamp were taken through a 1'' slit the day after each set of observations. Given the position of the comparison stars on the chip, the wavelength range 780 to 1000 nm was used in this analysis. The CCD response governs the red edge of this range, and the spectral response is similar to that of the MEarth bandpass. 

After bias subtraction and flat fielding, we used the comparison stars to correct for the time varying zeropoint of the system. For each exposure, we extracted 1D spectra from the images using the optimal extraction of \citet{horne.1986.oeas}, and divided the total flux (summed over wavelengths) of GJ1214 by the total flux in all the comparison stars. Theoretical error bars calculated from photon statistics alone were assigned to each point. Each exposure yielded $1-3\times10^8$ photons from GJ1214 and twice as many from comparison stars. 

The corrected GJ1214 light curves exhibit systematic trends which we correct for by fitting a second-order polynomial function of time. In \S\ref{Transits} we propagate the uncertainty from the systematics corrections through to the transit parameters. We searched for correlations between the relative flux and airmass, seeing, and positional shifts in the dispersion and cross-dispersion directions. The relationships were more complicated than low-order polynomials, so we did not attempt to remove them using common decorrelation techniques \cite[e.g.][]{burke.2010.notjx}.

We also tested whether the observed drifts in flux could be caused by a changing color-dependence of the atmospheric extinction along the line of sight. To do so, we applied differential photometry corrections to individual spectral channels {\em before} combining them, to allow each wavelength its own extinction. This procedure did not remove the systematic trends.

We suggest the following as a more probable explanation for the systematics. \citet{moehler.2010.cfrfscsuavd} found that the linear atmospheric dispersion corrector (LADC) on the telescope has surface features that affect is sensitivity across the field of view. Because the LADC is positioned before the field rotator in the optical path and rotates relative to the sky, individual stars can drift across these features and encounter throughput variations that are not seen by the other comparison stars. No rotationally-dependent flat-fields were applied to these data, although \citet{moehler.2010.cfrfscsuavd} provide a route to a possible correction.

The first two ``white'' light curves, normalized to their median out-of-transit flux level, are published in Table~\ref{tab-LC} and shown in Fig.~\ref{fig-vlt}.

A third transit observation was attempted on 2010 Jul 22. The brightest comparison star could not be used because it saturated mid-transit. The exposure times were cut in half immediately after egress, and a notable offset is visible in the transit light curve, perhaps due to an uncorrected non-linearity in the detector. Although the light curves of the first two transits were robust to the choice of comparison stars, the third changed significantly depending on which set of comparison stars was used. We show this transit in Fig.~\ref{fig-vlt}, but exclude it from all following analyses.

\begin{deluxetable}{ccc}
\tabletypesize{\normalsize}
\tablecaption{\label{tab-vi} Photometry of GJ1214}
\tablecolumns{3}
\tablehead{\colhead{Filter} & \colhead{Magnitude} & \colhead{Source}}
\startdata

$V $ & $14.71 \pm 0.03$ & this work\\
${I}$ & $11.52 \pm 0.03$ & this work\\
${J}$ & $9.750 \pm 0.024$  & 2MASS\tablenotemark{a}\\
${H}$ & $9.094 \pm 0.024$ & 2MASS\tablenotemark{a}\\
${K}$ & $8.782 \pm 0.024$ & 2MASS\tablenotemark{a}\\
\enddata
\tablenotetext{a}{\cite{skrutskie.2006.ms2}}

\end{deluxetable}

\section{Rotation Period of GJ1214}
\label{Rotation}
With a growing understanding of the systematic effects present in MEarth data, we revisit the issue of GJ1214's intrinsic variability. In the discovery paper for GJ1214b \citep{charbonneau.2009.stnls}, we stated that the dominant periodicity seen in the out-of-transit light curve of the star GJ1214 had an 83 day period, implying that this was the rotation period of the star. Here, we { revisit the question of GJ1214's rotation period with another season of observations.}

Semi-stable spot complexes on the surface of a star imprint photometric modulations that can be approximated as a sinusoid with a fundamental period that matches the stellar rotation period. { We search each year's light curve with a weighted, least-squares periodogram that has been modified to simultaneously fit for stellar variability along with scaled templates of known systematic effects. These systematics are discussed in the next several subsections. To account for the likely evolution of spots with time, we investigate the 2008, 2009, 2010 data sets separately and do not require a coherent sine curve to persist over multiple years' data. }

\subsection{Avoiding Persistence in MEarth Light Curves}

The MEarth detectors are subject to image persistence; pixels that are illuminated in one exposure can show enhanced dark current in subsequent exposures, which decays exponentially with a half hour time scale. Because the dark current in a given pixel depends on how recently that pixel was illuminated, differential photometry light curves can show baseline shifts between observations taken at different cadences, as well as `ramps' at the start of a high-cadence sequence of exposures.

Correcting for these changing baseline shifts would require a simultaneous modeling of the complete photon detection history of every pixel and is impractical. During MEarth's normal survey mode, we purposely center subsequent targets on different pixels to avoid persistent charge stacking up. As the effect is most noticeable for data with cadence shorter than 5 minutes, we circumvent the problem by throwing out from the rotation period analysis all but the first point of any segment with such cadence.

 \begin{figure}
\centering
\includegraphics[width=3.5in, clip=true]{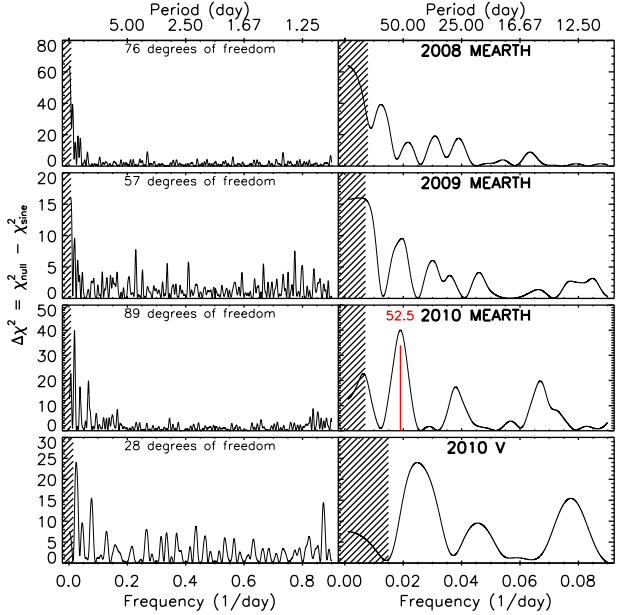}
\caption{Periodograms showing the $\Delta \chi^2$ improvement achieved by fitting a sinusoid of a given period + common mode + meridian flip over a null model consisting only of common mode + meridian flip for MEarth ({\em top 3 panels}) and FLWO V-band photometric monitoring ({\em bottom}). Possible periods as short as 1 day ({\em left}) and a zoom in to longer rotational periods ({\em right}) are shown. Periods for which less than one full cycle is observed per season are shaded.}
\label{fig-periodograms}
\end{figure}

\begin{figure}
\centering
\includegraphics[width=3.5in]{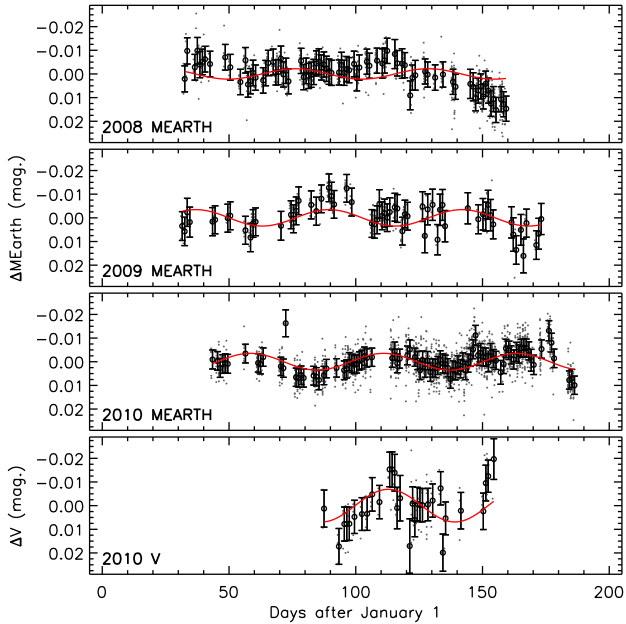}
\caption{{ Out-of-transit light curves} of GJ1214 from the MEarth Observatory ({\em top three panels}) and in V from the FLWO 1.2m ({\em bottom panel}). Individual exposures ({\em gray points}) and nightly binned values with errors that include the systematic `jitter'  ({\em black circles}, see text for details) are shown for each. A sine curve at the proposed 53 day period (derived from 2010 MEarth data) is shown ({\em solid lines}) at the best-fit phase and semiamplitude for each light curve. The MEarth points have been corrected for the best-fit common mode and meridian flip decorrelation.}
\label{fig-rotation}
\end{figure}

\subsection{Adding a Systematic `Jitter'}

Even having removed the highest cadence data, the sampling in the MEarth GJ1214 light curve can vary from N=1 to N=35 points per night, with a typical theoretical noise limit per point of 3 millimagnitudes. In a strict least-squares sense, if the noise in our data were accurately described by an uncorrelated Gaussian process (e.g. photon noise), then every exposure should be allowed to contribute on its own to the period search, meaning the uncertainty associated with each night would go down as $1/\sqrt{N}$. Although MEarth telescopes have achieved such white noise down to millimagnitude levels within  individual nights \citep{charbonneau.2009.stnls}, photometric variations between nights are most likely dominated by subtle changes in the telescope that are not corrected by our calibration efforts. The $1/\sqrt{N}$ weighting scheme would unfairly bias a period search to fit only a few well-sampled nights.

To account for this in each light curve, we remove all in-transit exposures and bin the data to a nightly time scale. For each night with $N$ data points, we calculate an inverse variance weighted mean flux and time, using the theoretical errors calculated for each exposure in the weighting. To each of these nightly bins we assign an error given by
\begin{equation}
\sigma_{nightly} = \sqrt{\sigma_{bin}^2 + \sigma_{jitter}^2}
\end{equation}
where $\sigma_{bin}$ is the intrinsic standard error on the mean of the nightly bin ($RMS/\sqrt{N-1}$) and $\sigma_{jitter}$ is a constant noise floor term to capture the night-to-night calibration uncertainty.

The signal lost in the binning process should be minimal. Preliminary searches of unbinned data and visual inspection of high cadence nights revealed no significant periodic signal at periods shorter than 1 day. { Under the (untested) assumption that the stellar spin is roughly aligned with the orbital angular momentum, the upper limit on the projected rotation velocity of $v\sin i < 2$ km s$^{-1}$ would correspond to a rotation period $P_{rot} > 5$ days.}

{ We use observations on successive nights to estimate $\sigma_{jitter}$ for the MEarth and the V-band light curves for each observing season. This assumes, on the basis of the apparent lack of short term variability, that the flux difference between pairs of nights is dominated by systematics. We find $\sigma_{jitter}$ = 0.0052, 0.0058, 0.0038, 0.0067 magnitudes for the 2008, 2009, 2010 MEarth and 2010 V light curves. }As these values are comparable to the predicted noise for most exposures, the quadrature addition of $\sigma_{jitter}$ means we weight most nights roughly equally. 

\subsection{Correcting for Meridian Flips in MEarth Light Curves}
MEarth's German Equatorial mounts rotate the detectors $180^\circ$ relative to the sky when switching from negative to positive hour angles. Thus, the target and comparison stars sample two different regions of the detector. Given imperfect flat field corrections, an offset is apparent in many MEarth light curves between exposures taken on either side of the meridian. 

To account for this effect, we allow different sides of the meridian to have different zeropoints. We construct a meridian flip template $m(t_i)$, which for an unbinned light curve would consist of binary values corresponding to the side of the meridian at each time stamp $t_i$. By extension, for each night of the binned data $t_i$, we define
\begin{equation}
m(t_i) = n_+/(n_+ + n_-)
\end{equation}
where $n_+$ and $n_-$ are the number of data points with positive or negative hour angles in a given nightly bin. We allow a scaled version of this template to be fit simultaneously with the period search.

\subsection{Correcting for Water Vapor in MEarth Light Curves}

Because the wide MEarth bandpass overlaps significant water absorption features in the telluric spectrum, the color-dependence of the throughput of our observing system is sensitive to the precipitable water vapor (PWV) in the column overhead. The fraction of stellar photons lost to water vapor absorption from a typical MEarth target M dwarf is much larger than the fraction lost from the (typically solar-type) comparison stars. When PWV along the line of sight to a star varies, a crucial assumption of simple differential photometry - that stars are experiencing the same losses - is violated. 

Although the variations in any particular light curve might come either from the PWV induced noise or from intrinsic stellar variability, we can harness the ensemble of M dwarfs observed by MEarth at any particular time to characterize and correct for this effect. To do so, we construct a ``common mode'' template by robustly (median) binning all the differential photometry light curves of all M dwarfs observed on all eight MEarth telescopes into half hour bins. This averages out uncorrelated stellar variability and serves as an estimate of the atmospheric variation that is common to all red stars observed at a given time.  We only use data during times when we have $>50$ and $>30$ targets contributing to a bin. The strongest periodicities in the common mode templates are 25.1 days for spring 2009 and 14.5 days for spring 2010 \citep{irwin.2010.amefsrpfmfmts}. If left uncorrected, such periodicities could appear as spurious intrinsic stellar variability.

To correct the GJ1214 light curve for this effect, we interpolate the ``common mode'' to the unbinned time stamps. We then perform the nightly binning on it to construct a common mode template $c(t_i)$, which we use for simultaneous decorrelation (see next section). This binning is justified because the typical common mode variation within a night is typically at the level of 1-2 millimagnitudes, much smaller than the night-to-night or week-to-week changes we hope to correct.

\subsection{Periodograms}

We generate a periodogram by calculating the $\chi^2$ of a weighted  linear fit of the light curve $\Delta F(t_i)$ (in magnitudes) to a model
\begin{equation}
\Delta F_{sine}(t_i) = A\sin\left(\frac{2\pi(t_i - t_0)}{P_{rot}}\right) + B\ m(t_i) + C \ c(t_i) + D
\label{eq-sine}
\end{equation}
where A is the semiamplitude of the sinusoidal variability, $t_0$ is an epoch, $P_{rot}$ is the stellar rotation period, B and C are scale factors for the systematics, and D is a constant offset. We compare this $\chi^2$ to that of the null hypothesis that $\Delta F$ is explained by the systematics alone
\begin{equation}
\Delta F_{null}(t_i) = B\ m(t_i) + C\ c(t_i) + D.
\label{eq-null}
\end{equation}
Mathematically, this procedure would be identical to traditional least-squares periodograms \citep{lomb.1976.lfausd,scargle.1982.satsasasausd} if we fixed $B=C=0$. In \citet{irwin.2010.amefsrpfmfmts}, we use a similar method to estimate photometric rotation periods for a sample of 41 MEarth M dwarf targets and test its sensitivity with simulations.

In Fig.~\ref{fig-periodograms} we plot the $\chi^2$ improvement ($\Delta \chi^2 = \chi^2_{\rm null} - \chi^2_{sine}$) between these two hypotheses for each of the three MEarth seasons and the short 2010 V-band campaign. { Periods for which less than one cycle would be visible have been masked.

The most prominent peak among all the periodograms in Fig.~\ref{fig-periodograms} is that at $53$ days from the 2010 MEarth data, which corresponds to a semiamplitude of $A=3.5\pm0.7$ millimagnitudes, where the uncertainty has been estimated from the covariance matrix of the linear fit. We estimate the false alarm probability (FAP) for this period by running the complete period search on $10^4$ time series that consist of the best-fit scaled versions of $m(t_i)$ and $c(t_i)$ and randomly generated Gaussian noise set by $\sigma_{nightly}$, recording the $\Delta \chi^2$ of the best peak from each iteration. We find that a FAP of $10^{-4}$ corresponds to $\Delta \chi^2=28$, much less than the achieved $\Delta \chi^2 =41$.

There is a nearby, but statistically insignificant, peak at 51 days in the 2009 MEarth data. Both 2008 and 2009 MEarth light curves are dominated by long-period trends that are unresolvable in each year. Of the resolved peaks in the 2008, the strongest is at 81 days (FAP $< 10^{-4}$). In spite of the formal significance of this last peak, we caution that mid-season changes to the then still uncommissioned observatory might also account for the variations seen. One conclusion is robust; our 3 years of MEarth light curves show no evidence for any rotational modulation with a period shorter than 25 days.

Given that 2010 had the most uniform sampling and cadence, we tentatively suggest $P_{rot} = 53$ days as our current best estimate of GJ1214's likely rotation period. Fig.~\ref{fig-rotation} shows the 3 MEarth and 1 V binned light curves with a sinusoid whose period has been fixed to our estimated $P_{rot}=53$ days but whose amplitude and phase have been fitted to the data. The fit is acceptable for 2009 MEarth, but clearly fails for the 2008 MEarth data.

We stress the caveat that the true rotation period could instead be a longer multiple of our quoted 53 day period (e.g. $P_{rot} \approx 100$ days) if the star exhibits multiple, well-spaced active regions. This kind of harmonic confusion appeared and was addressed in studies of Proxima Cen \citep{benedict.1998.ppcbsuhstfgsspv, kiraga.2007.ards}. Preliminary data for GJ1214 collected in 2011 while this paper was under review do not seem to show evidence for a 53 day period, preferring instead a much longer one. Due to this factor-of-n uncertainty in the true rotation period, we do not quote a formal error bar on our 53 day period estimate.}

\begin{figure}
\centering
\includegraphics[width=3.5in]{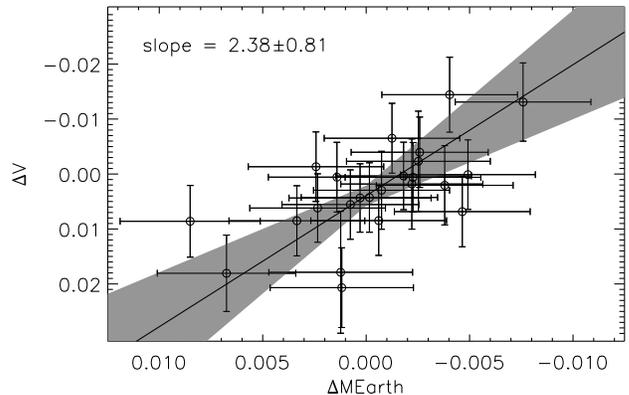}
\caption{For a sub-sample of nights measured in both MEarth and V bandpasses, the nightly bins plotted against each other with the assigned $\sigma_{bin}$ error bars. The best-fit slope ({\em black line}) and 1$\sigma$ interval ({\em shaded region}) are shown.}
\label{fig-mearth_v}
\end{figure}

\subsection{Chromatic Spot Variation}
{ 
By itself, the V-band light curve prefers a period of 41 days (Fig.~\ref{fig-periodograms}; FAP = 5$\times 10^-4$). When forced to fit a 53 day period (Fig.~\ref{fig-rotation}), these data show a phase offset of only 2 days relative to the simultaneous MEarth data. The semiamplitude of this fit is $7 \pm 3$ millimagnitudes, twice that seen in the MEarth bandpass.

There are 28 nights when observations were obtained in both MEarth and V band. In Fig.~\ref{fig-mearth_v} we plot the nightly bins against each other; the apparent correlation suggests that the two instruments are observing the same stellar variability across two bands and not telescope systematics. We fit a line to the relation, accounting for errors in both $\Delta$MEarth and $\Delta$V \citep{press.2002.nrsc} and find a slope of 2.4$\pm$0.8. While the significance is marginal, we take this as further evidence that the amplitude of the variability in V is greater than that in the MEarth band. 

If starspots have a temperature ($T_{\bullet}$) that is only modestly lower than the stellar effective temperature ($T_{\rm eff}$), the color-dependence will arise from the spectrum of the spot rotating in and out of view. The factor of 2 we see would be consistent with $T_{\bullet}/T_{\rm eff}\approx90-95\%$ as is commonly assumed in M dwarf eclipsing binaries \citep[e.g.][]{morales.2009.aplebd, irwin.2009.3bvmebsdmo}. Totally dark spots ($T_{\bullet} = 0$K) would produce less of a chromatic variation, but would still be sensitive to the spectral signature of the stellar limb-darkening \cite[e.g.][]{poe.1985.saatinbswlc}.}

\begin{deluxetable}{cclrr}
\tabletypesize{\normalsize}
\tablecaption{\label{tab-transits} Summary of Transit Light Curves}
\tablecolumns{5}
\tablehead{\colhead{Cycle} & \colhead{UT Date} & \colhead{Instrument} & \colhead{RMS (ppm)} &\colhead{Cadence (sec.)}}
\startdata

0	&	2009 May 13	&	MEarth	&	2950		&	60\\
9	&	2009 May 29	&	FLWO	&	1960		&	45\\
9	&	2009 May 29	&	MEarth	&	1580		&	(binned) 45 \\
11	&	2009 Jun 01	&	FLWO	&	2060		&	45\\
11	&	2009 Jun 01	&	MEarth	&	1240		&	(binned) 45 \\
21	&	2009 Jun 17	&	MEarth	&	1620		&	(binned) 45\\
221	&	2010 Apr 29	&	VLT		&	380		&	72		\\
228	&	2010 May 10	&	MEarth	&	1770		&	(binned) 45 \\
233	&	2010 May 18	&	VLT		&	350			&	72\\

\enddata
\end{deluxetable}

\section{Fitting the Transit Light Curves}
\label{Transits}
\begin{figure*}
\centering
\includegraphics[width=7in]{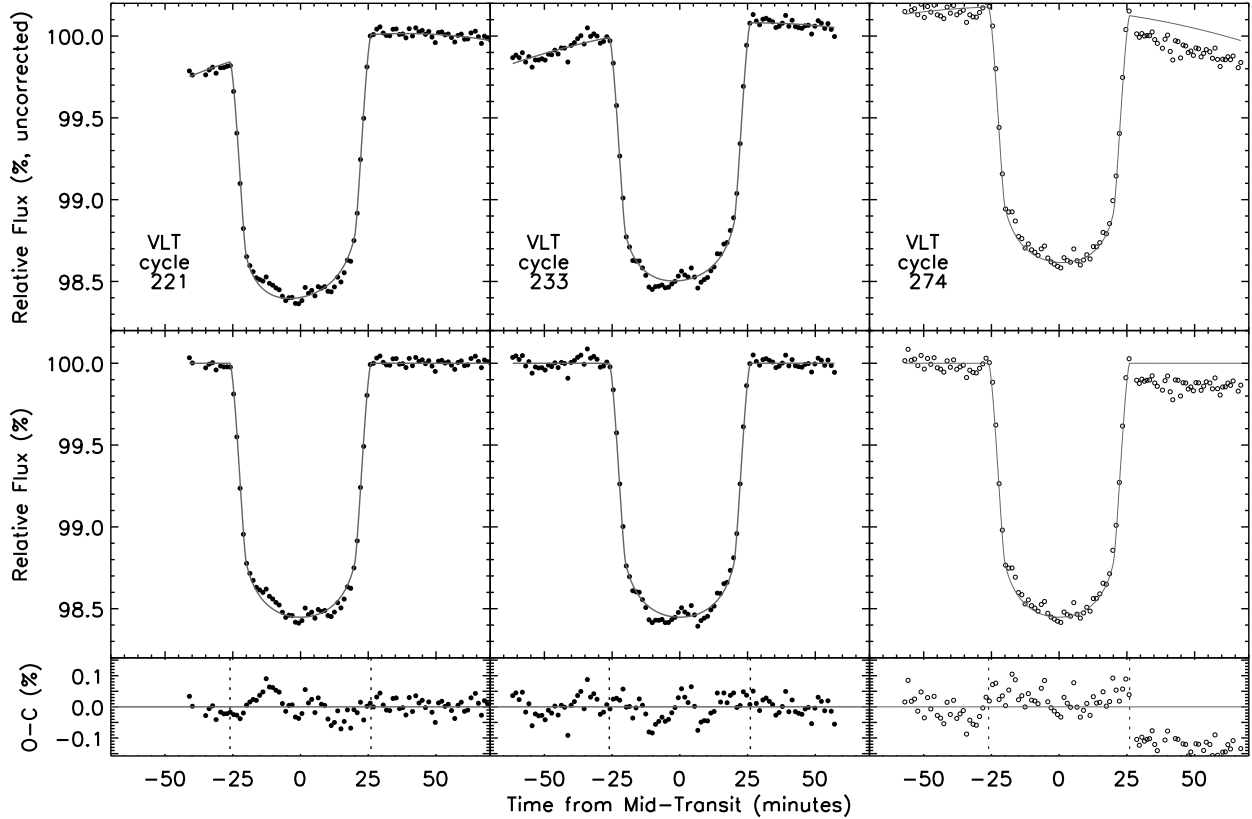}
\caption{Transit light curves from summed FORS spectra from the VLT, before ({\em top}) and after ({\em middle}) subtracting a second-order polynomial function of time. The third transit ({\em open circles}) exhibited larger uncorrectable systematics than the first two (see text), and was excluded from further analysis. Residuals from the quoted model ({\em gray lines}) are also shown ({\em bottom}), with dotted lines indicating the in-transit duration. }
\label{fig-vlt}
\end{figure*}

\begin{figure*}
\centering
\includegraphics[width=7in]{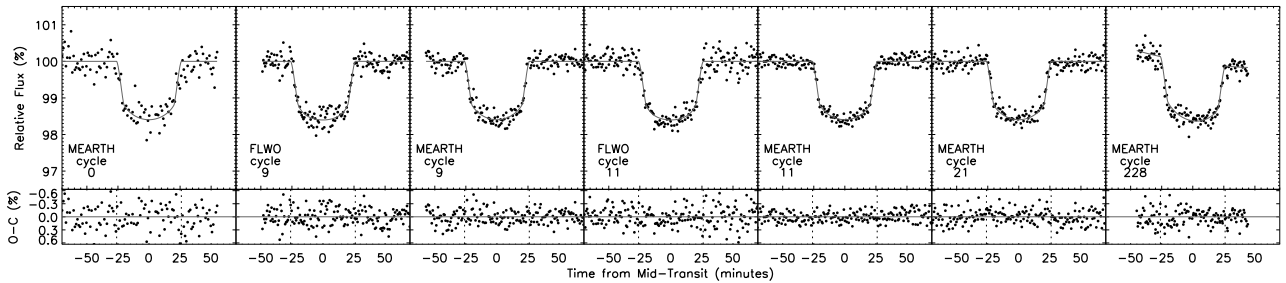}
\caption{Transit light curves from MEarth and KeplerCam. Residuals from the quoted model ({\em gray lines}) are also shown ({\em bottom}), with dotted lines indicating the in-transit duration. The cycle 228 MEarth transit shows a linear trend with airmass that is included in the fit. }
\label{fig-other}
\end{figure*}

\begin{figure*}
\centering
\includegraphics[width=6in]{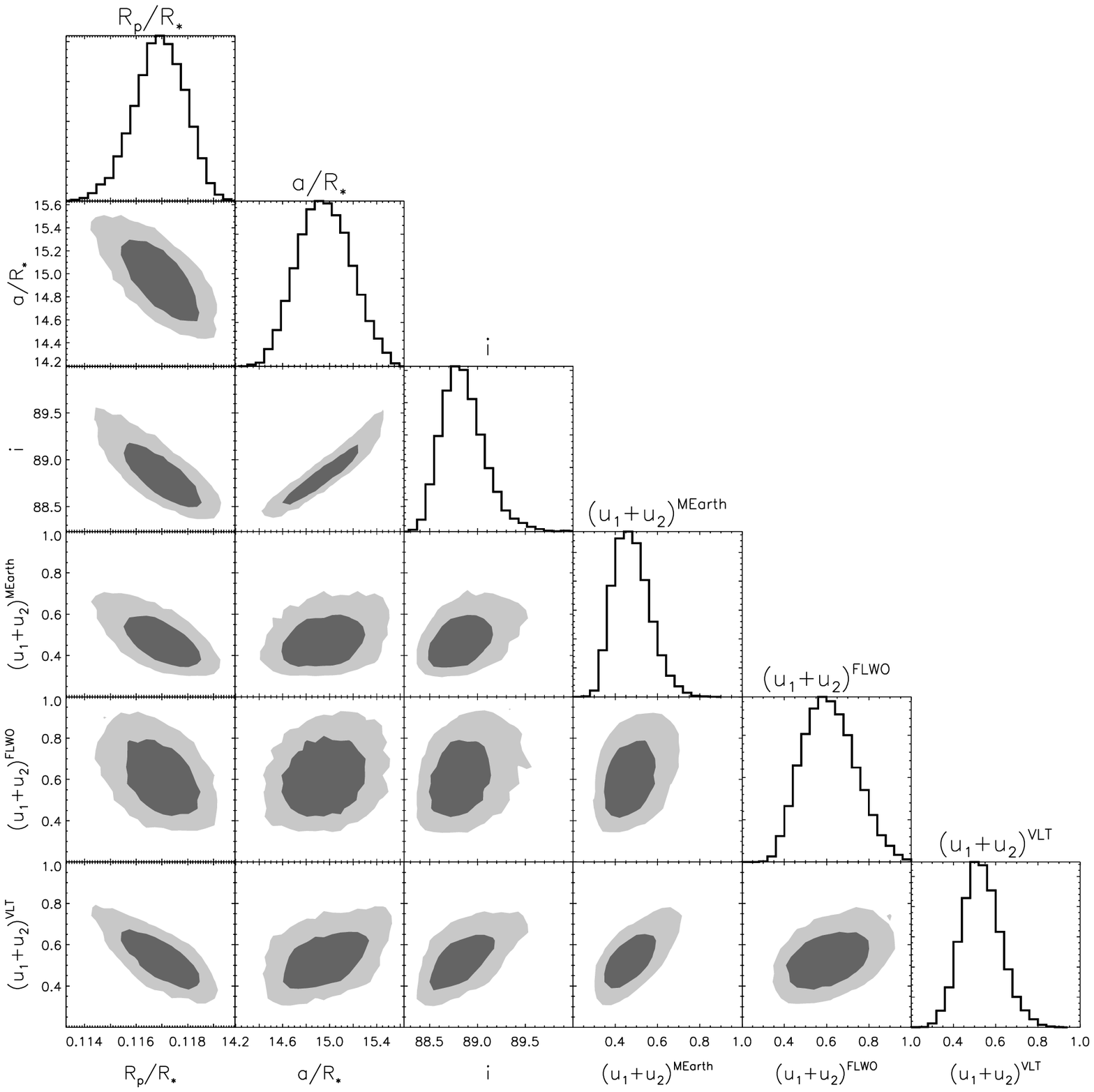}
\caption{Matrix showing the correlations between selected transit parameters from a simultaneous residual permutation analysis of 5 MEarth, 2 FLWO, and 2 VLT light curves. Histograms are shown for each parameter, as well as contours that contain $68\%$ ({\em dark gray}) and $95\%$ ({\em light gray}) of the samples for each pair of parameters. For each telescope system, we show a linear combination of the limb-darkening coefficients $u_1 + u_2$ that is strongly correlated with the other parameters.}
\label{fig-matrix}
\end{figure*}

We perform a simultaneous fit to the 4 transit light curves from MEarth and 2 from KeplerCam published by \citet{charbonneau.2009.stnls}, 1 more transit collected by MEarth in spring 2010, and 2 high-precision transits from the VLT. This totals 9 light curves of 7 independent transits, as shown in Figures~\ref{fig-vlt} and \ref{fig-other}. We employ a model corresponding to a circular planet transiting a smooth, limb-darkened star \citep{mandel.2002.alcpts} that has the following parameters: the planet-to-star radius ratio $R_{p}/R_{\star}$, the stellar radius $R_{\star}$, the total transit duration $t_{\rm 14}$, two quadratic stellar limb darkening parameters $u_1$ and $u_2$ for each of the 3 telescope systems used, and 7 mid-transit times $T_{c}$. The reparameterization of the scaled semimajor axis $a/R_{\star}$ and inclination $i$ in terms of $R_{\star}$ and $t_{\rm 14}$ substantially reduces the degeneracies in the problem, leading to an efficient exploration of the parameter space \citep{burke.2007.xtjmcpmb, carter.2008.aatlouc}.

We fix the orbital period to $P=1.5804043$ days \citep{sada.2010.rtseg}. { Given the upper limit on eccentricity from radial velocities and the short circularization time for GJ1214b \citep[$10^6$ years for $Q'_p = 100$ and $Q'_* = 10^6$, following][]{2008MNRAS.384..663R}, we assume an eccentricity $e=0$ throughout.} Where necessary to derive physical parameters from the geometric parameters in the light curve fit, we adopt the the stellar mass $M_{\star} = 0.157$ \msun~\citep{charbonneau.2009.stnls}; we describe how we propagate the $0.019$ \msun~ uncertainty on this value to the other errors in the next section.

To account for the systematic trends present in the VLT light curves, we introduce a correction to the baseline stellar flux parameterized as a parabola in time ($a + b t + c t^2$) to each transit. Most of the MEarth and FLWO light curves showed no significant systematic trends, so only a single out-of-transit baseline flux level was fit to each night. The MEarth light curve on the night of 2010 May 10 showed a strong correlation with airmass, so we also included a linear trend with airmass for this one night.

\subsection{$\chi^2$ minimization}
\label{fit}

We determine the best-fit values of the 30 model parameters by using an implementation of the Levenberg-Marquadt (LM) routine called MPFIT \citep{markwardt.2009.nlfwm} to minimize the value of
\begin{eqnarray}
\chi^2 &=& \displaystyle \sum\limits_{i=1}^{N}\frac{(F_i - F_{\rm model})^2}{\sigma_{\rm i}^2} 
\label{chi}
\end{eqnarray}
where $F_i$ are the $N=1495$ flux measurements (Table~\ref{tab-LC}), $\sigma_i$ are their uncertainties, $F_{\rm model}$ is the model described above.

After this initial fit, the uncertainty estimates for each light curve were increased until the reduced $\chi^2$ of the out-of-transit residuals was unity and the fit was repeated.

While the LM fit provides a linearized estimate of the covariance matrix and errors of the parameters, this estimate is too precise because the method a) does not fully sample along non-linear correlations between highly-degenerate parameters and b) does not account for correlations between data points. In what follows, we calculate more conservative and realistic errors through a bootstrap method that addresses both these issues.

\subsection{Error Estimates by Residual Permutation}
\label{prayer}
Although the noise in the MEarth and FLWO light curves is well described by a white Gaussian process, the much lower photon noise in the VLT light curves reveals underlying low level serial correlations among data points, or `red' noise. The autocorrelation function of the VLT residuals is above 0.25 out to 4 and 2 data point lags respectively for the two transits, and binning the residuals by N points reduces the scatter more slowly than $1/\sqrt{N}$. To quantify the excess uncertainty in the fitted parameters due to this red noise, we perform a `residual permutation' bootstrap simulation \citep{moutou.2004.armteo,gillon.2007.asirmn4,desert.2009.scmate1} that fits resampled data while preserving the correlations between data points. This analysis is carried out simultaneously for the VLT, FLWO and MEarth light curves. 

After subtracting the best-fit model from the ensemble of light curves, we perform $2\times10^4$ iterations of the following procedure. Preserving the time stamps for all the exposures, we cyclicly permute the residuals for each light curve by a random integer (shifting along the series and wrapping back from the last exposure to the first), inject the best-fit model back into the set of shifted residuals, perform the LM fit on the simulated light curve ensemble, and record the results. To initialize the parameters for the LM fit, we select guesses drawn from a multivariate Gaussian distribution whose covariance matrix is scaled up by $2^2$ (i.e. $2 \sigma$) from the LM's estimate. We excised those bootstrap samples that found a best fit with unphysical limb-darkening parameters \citep[$u_1 <  0, u_1 + u_2 > 1, u_1 + 2u_2 < 0$,][]{burke.2007.xtjmcpmb}.

\begin{deluxetable*}{lcc}
\tabletypesize{\normalsize}
\tablecaption{\label{tab-params} GJ1214b Parameters and Uncertainties}
\tablecolumns{3}
\tablehead{\colhead{Parameter} & \colhead{Previous Work} & \colhead{This Work\tablenotemark{a}}}
\startdata

$R_p/R_*$  &$ 0.1162 \pm 0.0007 $& $0.1171\pm0.0010$ \\
$a/R_{*}$ &$ 14.66 \pm 0.41$ &$14.93\pm0.24$ \\
$i$ (deg.) &$88.62^{+0.35}_{-0.28}$& $88.80^{+0.25}_{-0.20}$ \\
$b$ & $0.354^{+0.061}_{-0.082}$& $0.313^{+0.046}_{-0.061}$ \\
$\rho_* (g/cm^3)$ & $23.9\pm2.1$ & $25.2\pm1.2$ \\
$R_* (R_\sun)$\tablenotemark{b} &$ 0.2110 \pm 0.0097$ & $0.2064^{+0.0086}_{-0.0096}$ \\
$R_p (R_\earth)$\tablenotemark{b}  & $2.68 \pm 0.13 $& $2.64\pm0.13$ \\
$t_{12} = t_{34}$ (minutes)\tablenotemark{c} & \nodata& 
$6.00^{+0.24}_{-0.25}$ \\
$t_{23}$ (minutes)\tablenotemark{c} & \nodata & $40.10^{+0.46}_{-0.43}$ \\
$t_{14}$ (minutes)\tablenotemark{c} & \nodata & $52.11^{+0.25}_{-0.22}$ \\

$P$ (days) 	& $1.5804043 \pm 0.0000005$ &   $ 1.58040490 \pm 0.00000033$\\
$T_0$ (${\rm BJD_{TDB}}$) & $2454966.52506 \pm  0.00006$\tablenotemark{d} & $2454966.525042 \pm 0.000065$

\\
\enddata
\tablenotetext{a}{Best-fit value and confidence intervals that exclude the upper and lower 15.9\% of the residual permutation bootstrap samples. Where upper and lower limits differ by $>10\%$, asymmetric errors are shown.}
\tablenotetext{b}{Uncertainty calculated as described in the text assuming $M_{\star} = 0.157\pm0.019$\msun}.
\tablenotetext{c}{Times between contact are $t_{12} = t_{34}$ = ingress/egress time, $t_{23}$ = duration of full eclipse, $t_{14}$ = total duration of transit. $t_{14}$ was fit directly from the light curve, and the others were calcuated from the fitted parameters \citep[see][]{seager.2003.uspspfeptlc, carter.2008.aatlouc}.}
\tablenotetext{d}{Converted from $BJD_{UTC}$ to $BJD_{TDB}$ and added 1 cycle for ease of comparison.}
\tablerefs{\citet{charbonneau.2009.stnls}, \citet{sada.2010.rtseg}}
\end{deluxetable*}

\begin{deluxetable}{lrrr}
\tabletypesize{\normalsize}
\tablecaption{\label{tab-ld} Inferred GJ1214 Limb-darkening Coefficients}
\tablecolumns{4}
\tablehead{\colhead{Instrument} & \colhead{$\lambda$ range (nm)} & \colhead{$u_1$\tablenotemark{a}} & \colhead{$u_2$\tablenotemark{a}}}
\startdata

MEarth & 715-1000 &$0.53\pm0.13$ & $-0.08\pm0.21$\\
VLT-FORS2 & 780-1000 & $0.34\pm0.31$ & $0.28\pm0.46$\\
KeplerCam $z$ & 850-1000& $0.26\pm0.11$ & $0.26\pm0.19$\\

\enddata
\tablenotetext{a}{Best-fit value and confidence intervals that exclude the upper and lower 15.9\% of the residual permutation bootstrap samples.}
\end{deluxetable}

In Table~\ref{tab-params}, we quote the best-fit value for each parameter and uncertainty bars that exclude the lower and upper 15.9\% of the bootstrap samples (i.e. the central 68.3\% confidence interval), where parameters that were not directly fit have been calculated analytically from those that were \citep{seager.2003.uspspfeptlc}. Having fit quadratic limb-darkening parameters for each instrument, we present similar confidence intervals of the coefficients $u_1$ and $u_2$ in Table~\ref{tab-ld}.

Fig.~\ref{fig-matrix} summarizes the correlations among the parameters $R_p/R_{\star}$, $a/R_{\star}$, $i$, and the linear combination of the limb-darkening parameters $u_1 + u_2$ for the three telescope systems. As the difference between central and limb specific intensities, $u_1+u_2$ correlates strongly with the transit depth and $R_p/R_{\star}$. We also show in Fig~\ref{fig-matrix} the bootstrap histograms for these parameters.

Although we use $R_{\star}$ as a fit parameter, the quantity that is actually constrained by the light curves is $\rho_{\star}$. To calculate the true uncertainty on $R_{\star}$ (and $R_p$), we calculate $\rho_{\star}$ for all of our bootstrap samples using the fixed $M_{\star}=0.157$\msun, assign values of $M_{\star}$ drawn from the appropriate Gaussian distribution, and recalculate $R_{\star}$ from $\rho_{\star}$ and $M_{\star}$. 

We find results consistent with \citet{charbonneau.2009.stnls}. Despite the high precision of the VLT light curves, the uncertainties for most parameters are comparable to the earlier work, and that on $R_p/R_{\star}$ is slightly larger. This is due in part to the correlated noise analysis we perform that \citet{charbonneau.2009.stnls} did not. More significantly, \citet{charbonneau.2009.stnls} fixed the quadratic limb-darkening parameters to theoretical values while we fit for them directly.  The extra degrees of freedom allowed by our relaxation of astrophysical assumptions are known to increase the uncertainty on $R_p/R_{\star}$ \citep{burke.2007.xtjmcpmb,southworth.2008.hstepila}. { For comparison, \citet{charbonneau.2009.stnls} used coefficients appropriate for a 3000K, $\log g = 5$, PHOENIX atmosphere, specifically Cousins $I$ \citep[tabulated in][]{bessell.1990.up} coefficients \citep[$u_1 = 0.303, u_2=0.561$;][]{claret.1998.vmsnll} as an approximation for the MEarth bandpass and Sloan $z$ coefficients \cite[$u_1 = 0.114, u_2=0.693$;][]{claret.2004.nlsamisfc2tssg} for the KeplerCam FLWO data. While our individual fitted values differ from these, the integral over the stellar disk ($1-u_1/3-u_2/6$) is very well reproduced.}

\begin{deluxetable}{rlc}
\tabletypesize{\normalsize}
\tablecaption{\label{tab-times} GJ1214 Transit Times}
\tablecolumns{3}
\tablehead{\colhead{Cycle\tablenotemark{a}}& \colhead{$T_{c}$ (days)\tablenotemark{b}} & \colhead{$\sigma$ (days)}} 
\startdata
       0 & $2454966.525207$ & $0.000351$ \\ 
       9 & $2454980.748682$ & $0.000104$ \\ 
      11 & $2454983.909507$ & $0.000090$ \\ 
      21 & $2454999.713448$ & $0.000155$ \\ 
     221 & $2455315.794564$ & $0.000066$ \\ 
     228 & $2455326.857404$ & $0.000110$ \\ 
     233 & $2455334.759334$ & $0.000066$ \\ 
\enddata
\tablenotetext{a}{Time from the first MEarth transit, measured in units of the orbital period.}
\tablenotetext{b}{Mid-transit times $T_c$ are given as Barycentric Julian Dates in the Terrestrial Time system (${\rm BJD_{TDB}}$). To the best precision afforded by these transits, these times can be converted to a UTC based system via the approximation ${\rm BJD_{UTC}} \approx {\rm BJD_{TDB}} - (32.184 + N)/86400$ where N=34 is the number leap seconds between UTC and TAI at the time of these transits \citep{eastman.2010.abtmahbjd}.}
\end{deluxetable}

\subsection{Transit Timing Results}
\label{transit-ttv-results}
Mid-transit times (equivalent to times of inferior conjunction given the assumed circular orbit) are printed in Table~\ref{tab-times}. Our uncertainty estimate on the VLT transit times is $50\%$ larger than it would be if we had ignored correlations in the data, but still only 6 seconds. This uncertainty is within a factor of two of the uncertainty on the highest-precision transit times yet measured from either the ground or space: the 3-second measurements of HD189733b with {\em Spitzer} \citep{agol.2010.c1fftems}. For stars of comparable brightness to GJ1214, large aperture ground based telescopes offer a powerful tool for precision transit times.
 
\begin{figure}
\centering
\includegraphics[width=3.5in]{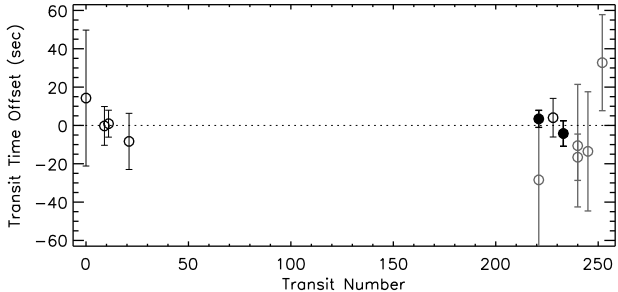}
\caption{Deviations in the times of transit from the new best linear ephemeris, including transits from the VLT ({\em filled black circles}), MEarth ({\em open black circles}), and work by \citet{sada.2010.rtseg} ({\em open grey circles}). }
\label{fig-times}
\end{figure}

A revised linear ephemeris is shown in Table~\ref{tab-params} derived with a weighted least squares  method from our transit times and those of \citet{sada.2010.rtseg}. Residuals are shown in Fig.~\ref{fig-times}. The reduced $\chi^2$ of the linear fit $\chi^2_{\nu}=0.62$ gives no indication of transit timing variations over the time scales probed, and a Bayesian model comparison test \citep[see][]{burke.2010.notjx} does not show significant evidence for a model with linear and hypothetical sinusoidal components over a purely linear ephemeris. We note that a hypothetical 1.0 \mearth~planet in a 2:1 mean-motion resonance with GJ1214b would introduce $\sim100$ second transit timing variations that would be easily detected in these data \cite[following][]{bean.2009.attc}.

\subsection{Occulted Spots}

Occulted spots will appear in the transit light curve as a bump lasting roughly as long it takes the planet to move a distance $2(R_p + R_{\bullet})$ across the spot, assuming a circular spot with radius $R_{\bullet}$. For small spots ($R_{\bullet} << R_p$), this is roughly the transit ingress/egress time, or 6 minutes in the case of GJ1214b. The amplitude of a spot crossing event is determined by the fractional deviation in surface brightness occulted by the planet from the star's mean and could in principle be comparable to the transit depth itself for totally dark spots larger than the planet. 

We see evidence for spot occultations in the first two VLT transits shown in Fig.~\ref{fig-vlt}, in which the residuals from a smooth model show a 0.1\% brightening at the start of the first transit and near the middle of the second transit. These features last 5-10 minutes, as expected, and persist with comparable amplitudes in light curves generated separately from the blue and red halves of the FORS spectra. Their presence is robust to choice of comparison stars.

For the level of precision afforded by current data, we treat these possible spot occultations as excess red noise with correlations on the scale of the 6 minute spot crossing time. As such, the residual permutation method (\S\ref{prayer}) accounts for the uncertainty introduced by these features, regardless of their physical interpretation. Spots that are not occulted will not be accounted for in these errors estimates, but we discuss their influence in $\S\ref{implications}$.

\subsection{Unocculted Spots}
\label{spots-results}

Unocculted spots will also have an effect on the planetary parameters. By diminishing the overall flux from the star while leaving the surface brightness along the transit chord unchanged, increasing the coverage of unocculted spots on the star will make transits deeper. This is opposite the effect of occulted spots, which tend to fill in transits and make them shallower. Because we must apply a nightly normalization to each light curve to avoid systematics and compare across telescopes, this depth change carries through to the implied $R_p/R_{\star}$. 

To test whether unocculted spots are biasing individual measurements of $R_p/R_{\star}$, we repeat the fit and uncertainty estimation described in \S\ref{fit} and \S\ref{prayer} but allow the 8 transit epochs to have different values of $R_p/R_{\star}$. As the visible spot coverage changes when the star rotates, we might detect changes in the inferred values of $R_p/R_{\star}$. For this experiment we are interested in relative changes among the $R_p/R_{\star}$ values, so we fix the other geometric parameters $R_{\star}$ and $t_{\rm 14}$ and the limb-darkening parameters to their best fit values to effectively collapse along those dominant degeneracies.

\begin{figure}
\centering
\includegraphics[width=3.5in]{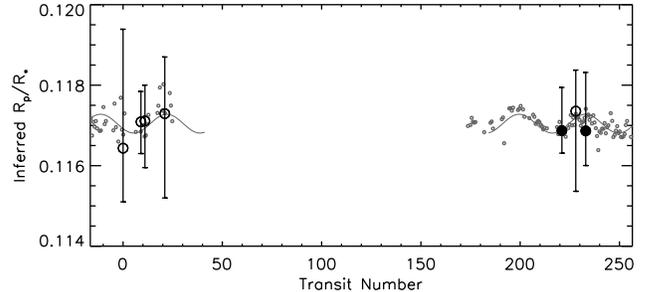}
\caption{Estimates of the apparent planet-to-star radius ratio at each epoch ({\em black circles} for MEarth/FLWO ({\em open symbols}) and VLT transits ({\em filled symbols})). Via the residual permutation estimate, the error bars include the uncertainty due to possible presence of occulted spots. The predicted variation in the apparent $R_{p}/R_{\star}$ due to the presence of unocculted spots (see \S\ref{implications}) is shown ({\em grey}), calculated directly from the nightly-binned MEarth photometry ({\em points}) or the best-fit sine curves to those data ({\em lines}).}
\label{fig-k}
\end{figure}

The inferred $R_p/R_{\star}$ for each transit is shown in Fig.~\ref{fig-k}, with uncertainties estimated from the residual permutation method. Because the other parameters were fixed in this fit, the ensemble of points are free to move up and down slightly on this plot; the uncertainties shown are more relevant to comparisons between epochs. In \S\ref{implications}, we will discuss our calculation of the predicted variations induced by unocculted spots consistent with the observed variability.

\section{Limits on Additional Transiting Planets}
\label{Search}
After clipping out known transits of GJ1214b, we investigate the light curve of GJ1214 for evidence of other transiting planets in the system. The light curve contains 3218 points spanning spring 2009 and 2010. 

To remove structured variability from the light curve before searching for transits, we employ an iterative filtering process that combines a 2-day smoothed median filter \citep{aigrain.2004.ppp} with a linear decorrelation against both external parameters (primarily the common mode, meridian flip, seeing, and pixel position) and light curves of other field stars \citep{kovacs.2005.tfawvs,tamuz.2005.cselplc,ofir.2010.sadclcwsusep}. Such filtering decreases the light curve RMS from 9.6 to 4.7 millimagnitudes per point.

Using a variant of the box-fitting least squares (BLS) algorithm \citep{kovacs.2002.baspt}, we search the filtered light curve for periodic rectangular pulses over a grid of period and transit phase. At each test period, we fix the transit duration to that of a mid-latitude transit of a circularly orbiting planet for GJ1214's estimated stellar mass (0.157 \msun) and radius (0.207 \rsun). { With the typical MEarth precision and cadence, our sensitivity depends only weakly on transit duration; violations of these assumptions will not substantially penalize our detection efficiency.}

At every grid point, we determine the best fit transit depth $D$ and the improvement $\Delta \chi^2$ over the $D=0$ null hypothesis, using a weighted least-squares method that includes red noise estimated from the light curve itself using the $\sigma_r$ formalism of \citet{pont.2006.enptd}. Following \citet{burke.2006.stepssilfswpoc1} we characterize candidates in terms of $\Delta \chi^2$ and $f=\max(\Delta\chi^2_{\rm each~transit}/\Delta\chi^2_{\rm total})$, which is the largest fraction that any one transit event contributes to the signal. 

\begin{figure*}
\centering
\includegraphics[width=7in]{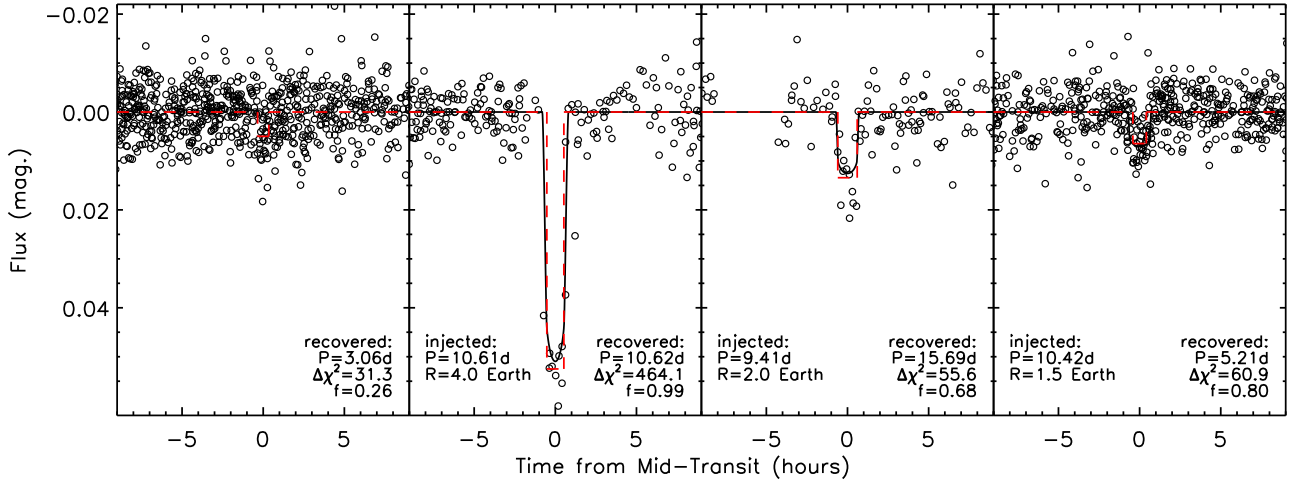} 
\caption{{\em Left:} The best candidate returned by the BLS search of the cleaned GJ1214 light curve. Bootstrap searches of simulated noise show the achieved $\Delta \chi^2$=31.3 has a large $20\%$ probability of false alarm. {\em Right three:} Examples of simulated transiting planets with periods near 10 days where the transits injected ({\em black lines}) into prefiltered light curves have crossed the detection threshold in the BLS search. All four panels show light curves phased to the period of the best fit BLS result ({\em red dashed lines}); the BLS does not necessarily recover the correct period.}
\label{fig-inject}
\end{figure*}

The best candidate found in the clipped GJ1214 light curve exhibited a $\Delta \chi^2$ of 31.3 and is shown in Fig.~\ref{fig-inject}. To estimate the significance of this value, we employ the bootstrap method of \citet{jenkins.2002.stecpdtp}. { Strictly speaking, the presence of remaining correlated noise in our filtered light curve means this method gives an overestimate of the false alarm probability, but the complicated correlation structure of the light curve make a more accurate significance estimate difficult to calculate. In practice, we have found the Jenkins method to provide an appropriate limit, even for light curves with substantial red noise.} We generate $10^3$ (so we can estimate the $\chi^2$ associated with a $1\%$ false alarm probability from 10 samples) transit-less light curves with Gaussian white noise and time sampling identical to that of the real light curve and performing the BLS search on these fake light curves. We find $20\%$ of them show values of $\chi^2 > 31.3$, suggesting our best candidate shown in Fig.~\ref{fig-inject} is not significant. 

These data place limits on the presence of other transiting planets in the GJ1214 system. Like \citet{burke.2006.stepssilfswpoc1}, \citet{croll.2007.ls1sstmsp} and \citet{ballard.2010.sapneoesg}, we simulate our sensitivity by injecting 8000 randomly phased, limb-darkened transits of 1.0, 1.5, 2.0, and 4.0 \rearth~planets with random periods. We then attempt to recover them with our transit search using objective detection criteria. To account for possible suppression from the filtering, we inject the transits into the raw light curves and reapply the filter at each iteration. We adopt two criteria for detection: $\Delta \chi^2>50$ (corresponding to a formal probability of false alarm of $10^{-5}$ by the bootstrap analysis) and $f<1.0$ (to ensure at least two events contribute). 

\begin{figure}
\centering
\includegraphics[width=3.5in]{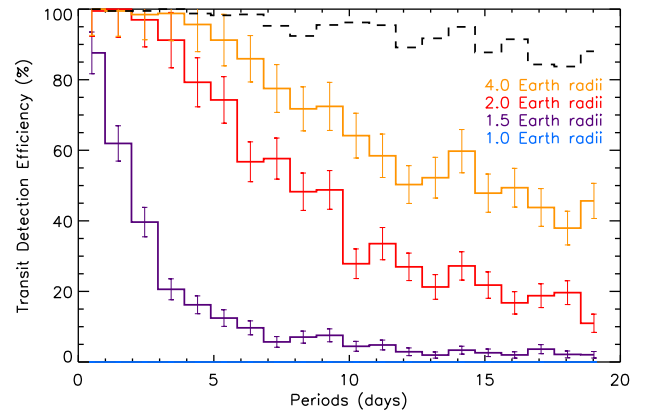}
\caption{The recovery fraction as a function of period for the labeled planetary radii ({\em solid, colored lines}) assuming the detection criteria $\Delta \chi^2>50$ and $f<1.0$. For the 4.0 \rearth~ case, we also show the recovery fraction after lifting the $f<1.0$ constraint ({\em black dashed line}); this is an estimate of the sensitivity to deep transits where only one event is necessary for a robust detection.}
\label{fig-sensitivity}
\end{figure}

To demonstrate this visually, we show in Fig.~\ref{fig-inject} examples of injected transits for each relevant radius, randomly selected from among the simulated planets with injected periods of $10 \pm 1$ days whose BLS results satisfied the detection criteria. Planets with periods near $10$ days are of particular interest; given GJ1214's low luminosity, they would be in the star's habitable zone. Fig.~\ref{fig-sensitivity} plots the fraction of injected planets that crossed the detection threshold as a function of period for each input planetary radius. The shape of the 4.0 \rearth~curve in  Fig.~\ref{fig-sensitivity} is driven largely by the $f<1.0$ criterion requiring multiple events are observed, whereas the 1.5 \rearth~curve is dominated by the need for sufficient in-transit S/N to get to $\Delta\chi^2 > 50$. With these criteria, our sensitivity falls below 50\% beyond periods of 15, 8, and 2 days for 4.0 \rearth, 2.0 \rearth, and 1.5 \rearth~ transiting planets. We had no sensitivity to 1.0 \rearth~ planets.

Fig.~\ref{fig-sensitivity} also shows the recovery fraction for 4.0 \rearth~planets without requirement that more than one event be observed. Neptune-sized transits of GJ1214 are so dramatic that they could be confidently identified by a single event (see Fig.~\ref{fig-inject}). Our simulations show that we are 90\% sensitive to transiting Neptunes around GJ1214 out to 10 days, and 80\% sensitive out 20 days. For smaller planets requiring multiple events, however, a significant volume of parameter space remains unconstrained.

\section{Discussion}
\label{Discussion}

\subsection{GJ1214 as a Spotted Star}

The slow rotation period we find implies a projected rotational velocity of 0.2 km s$^{-1}$ that is well below the $v \sin i \simeq 1$ km s$^{-1}$ detection limit of high-resolution rotation studies \citep{browning.2010.rmasm,delfosse.1998.rcafd,reiners.2007.nmlprcvsr,west.2009.flrild} but not inconsistent with long photometric periods detected for other field M dwarfs. \citet{benedict.1998.ppcbsuhstfgsspv} estimated a rotation period of 83 days for Proxima Cen, and recent photometric work with ASAS \citep{kiraga.2007.ards}, HATnet \citep{hartman.2009.pvsfdswh}, and MEarth \citep{irwin.2010.amefsrpfmfmts} has confirmed the presence of many field M dwarfs with $P_{rot}>10$ days.

Our $P_{rot}$ further implies that GJ1214 should exhibit signs of only weak magnetic activity \citep[e.g.][]{reiners.2007.fdmsmfvms, reiners.2007.nmlprcvsr}. Indeed, across three seasons of photometric monitoring, we see no evidence for flares in the MEarth bandpass, although their amplitude would be expected to be small in the near-IR. Activity induced chromospheric emission is not detected in either H$\alpha$ or the {Na I} D doublet in the HARPS spectra used to measure the radial velocities presented in  \citet{charbonneau.2009.stnls}. The relation between magnetic activity and kinematic age \citep{west.2008.carcssdsdrlsss} suggests that GJ1214 is $>3$ Gyr old.

We calculate GJ1214's $(U, V, W)$ space velocities \citep{johnson.1987.cgsvtuwaumg} in a left-handed system where $U > 0$ in the direction of the Galactic anti-center to be (-47,-4,-40) km s$^{-1}$.  These motions are consistent with membership in the Galactic old disk \citep{leggett.1992.icls}, lending further credence to an old age for GJ1214.

\subsection{Implications for Transmission Spectroscopy}
\label{implications}

When inferring the transmission spectrum of a planet, one hopes to attribute changes in the transit depth across different wavelengths to atmospheric absorption by the limb of the planet. If the transmission spectrum is sensitive to 5 scale heights ($H$) of the planetary atmosphere, the amplitude of the transit depth variations are $\Delta D_{\rm planet}(\lambda) = 10HR_p/R_{\star}^2$ or $0.001$ if GJ1214 has an hydrogen-rich atmosphere \citep{miller-ricci.2010.nats1}.  However, the presence of unocculted spots on the stellar surface can introduce transit depth variations $\Delta D_{\rm spots} (\lambda, t)$ that are a function of both wavelength and time. To aid ongoing and future work to study GJ1214b's atmosphere, we use a simple model to estimate the amplitude of the spot-induced contamination $\Delta D_{\rm spots} (\lambda, t)$.

We assume a fraction $s(t)$ of the star's Earth-facing hemisphere is covered with spots; $s(t)$ will change as the star rotates and the spots evolve. The observed out-of-transit spectrum $F_{\rm o.o.t.}(\lambda, t)$ is a weighted average of the spectrum of the unspotted photosphere $F_{\circ}(\lambda)$ and that of the presumably cooler spotted surface $F_{\bullet}(\lambda)$:
\begin{equation}
F_{\rm o.o.t.}(\lambda, t) = \left[1-s(t)\right]F_{\circ}(\lambda) + s(t)F_{\bullet}(\lambda).
\end{equation}
To simplify calculations, we neglect limb-darkening and treat each of the two components as having uniform surface brightness; tests with a limb-darkened spot model \citep{dorren.1987.fsmcss} indicate only pathological cases could change the following results by more than $10\%$.

When a planet with no atmosphere blocks light across a spot-free transit chord, it changes the relative weight of the two sources, causing the observed in-transit spectrum $F_{\rm i.t.}(\lambda, t)$ to shift away from that of the unspotted photosphere toward that of the spots:
\begin{eqnarray}
\frac{F_{\rm i.t.}(\lambda, t)}{F_{\rm o.o.t.}(\lambda, t)} &=& 1 - D(\lambda, t) \\
&=& 1- \left(\frac{R_p}{R_{\star}}\right)^2 - \Delta D_{\rm spots}(\lambda, t) \nonumber \\
&=& \frac{\left[1-s(t)-\left(\frac{R_p}{R_{\star}}\right)^2\right]F_{\circ}(\lambda) + s(t)F_{\bullet}(\lambda)}{ \left[1-s(t)\right]F_{\circ}(\lambda) + s(t)F_{\bullet}(\lambda)}. \nonumber
\end{eqnarray}
Making the assumption that the total fraction of flux lost to the presence of spots is small ($s(t)[1-F_{\bullet}(\lambda)/F_{\circ}(\lambda)] << 1$), we solve to find
\begin{equation}
\Delta D_{spots}(\lambda, t) \approx s(t)\left[1-\frac{F_{\bullet}(\lambda)}{F_{\circ}(\lambda)}\right]\times\left(\frac{R_p}{R_{\star}}\right)^2.
\label{eq-dspots}
\end{equation}
If the quantity $s(t)[1-F_{\bullet}(\lambda)/F_{\circ}(\lambda)]$ were not small, we probably would have observed larger amplitude and more frequent spot occultation events in the transit light curves in \S\ref{Transits}. We note the significant possibility that $s(t)$ never reaches 0; that is, there may persist a population of symmetrically distributed spots that never rotates out of view. Such an unchanging population could cause us to overestimate the true value of $R_p/R_{\star}$ by up to several percent.

To match our observations of GJ1214's variability ($\Delta F_{\rm o.o.t.}(\lambda, t)/\overline{F_{\rm o.o.t.}}$) to this model, we write $s(t) = \overline{s} + \Delta s(t)$ where $\overline{s}$ is the mean Earth-facing spot covering fraction and $\Delta s(t)$ can be positive or negative. With the same assumption as above, we find
\begin{equation}
\frac{\Delta F_{\rm o.o.t.}(\lambda, t)}{\overline{F_{\rm o.o.t.}}} \approx - \Delta s(t) \left( 1 - \frac{F_{\bullet}(\lambda)}{F_{\circ}(\lambda)}  \right).
\label{eq-var}
\end{equation}
In \S\ref{Rotation} we measured $\Delta F_{\rm o.o.t.}(\lambda, t)/\overline{F_{\rm o.o.t.}}$ to have a peak-to-peak amplitude of $1\%$ in the MEarth bandpass ($715 < \lambda < 1000$ nm). Eq.~\ref{eq-dspots} and~\ref{eq-var} are equivalent to assuming a value of $\alpha = -1$ in \citet{desert.2011.tse1isom}'s $\alpha f_{\lambda}$ formalism.

\citet{desert.2011.tse1isom} show that measurements that rely on comparing photometric transit depths across multiple transits could potentially mistake time-variability of $\Delta D_{\rm spots}(t)$ for a feature in the transmission spectrum. From Eq.~\ref{eq-dspots} and \ref{eq-var}, we estimate the peak-to-peak time-variability of $\Delta D_{\rm spots}(t)$ to have an amplitude of $0.0001$ in MEarth wavelengths over the rotation period of the star. This spot-induced variability is comparable to, but smaller than, the estimated uncertainty from the transits analyzed in this work and corresponds to an apparent change in planetary radius of $\Delta R_p = 70~{\rm km}$ or 1/2 the scale height of an  H$_2$-dominated atmosphere on GJ1214b \citep{miller-ricci.2010.nats1}. In Fig.~\ref{fig-k} we show the expected variation in the apparent planetary radius from unocculted spots and our individual $R_p/R_{\star}$ measurements. 

Using blackbody spectra for $F_{\bullet}(\lambda)$ and $F_{\circ}(\lambda)$, we can extrapolate with Eq.~\ref{eq-dspots} from the MEarth observations ($\lambda \approx 0.85 ~{\rm \mu m}$) to wavelengths accessible to {\em Warm Spitzer}. If we assume the spots are 300K cooler than the $T_{\rm eff}=3000$K stellar photosphere, we find $\Delta D(t)$ variability amplitudes of 0.00004 and 0.00003 in {\em Spitzer}'s 3.6 and 4.5 ${\rm \mu m}$ bandpasses. This allows robust comparison of transit depths between these wavelengths (D\'esert et al., in prep.). In the conservative limit that $T_{\bullet}=0$K, the variability amplitude would be achromatic.

Even when comparison across different epochs is unnecessary, as is the case for spectroscopic observations of individual transits \citep[e.g.][]{charbonneau.2002.depa,pont.2008.dahep0mts1wh}, unocculted spots can still introduce spurious wavelength features into the transmission spectrum. To place an upper limit on the amplitude of chromatic changes in $\Delta D_{\rm spots}(\lambda)$ within a given transit, we imagine an extreme scenario where $F_{\bullet}(\lambda)$ is identical to $F_{\circ}(\lambda)$ but with a very deep absorption line that does not appear in the unspotted spectrum, so the first factor in Eq.~\ref{eq-dspots} can vary with $\lambda$ between 0 and $s$. Making the fairly conservative assumption that the population of spots that rotates in and out of view is comparable to the symmetric population ($s < 4\max[\Delta s]$) we find that $\Delta D_{\rm spots}(\lambda) < 0.0003$ for GJ1214 at wavelengths near $1~{\rm \mu m}$. In practice, most features will show significantly lower amplitudes, although precise calculations of them and extrapolation to other wavelengths will require knowledge of the spot temperature and reliable model atmospheres in the 2500-3000K temperature regime. 

Given the sensitivity of current generation instruments, the known population of spots on GJ1214 do not pose a significant problem for ongoing studies of GJ1214b's atmosphere. Future transmission spectroscopy studies of GJ1214b, perhaps with the {\em James Webb Space Telescope}, comparing multiepoch, multiwavelength transit depths and aiming to reach a precision of $\sigma_{D(\lambda)} = 0.0001$  \citep[see][]{deming.2009.dctseuatsfjwst} will have to monitor and correct for the stellar variability. 

\subsection{Limiting Uncertainties of GJ1214b}

As shown in Fig.~\ref{fig-k}, stellar spots currently play a very small role in limiting our understanding of the bulk mass and radius of GJ1214b. Here we address what other factors provide the limiting uncertainties in the planet's physical parameters.

The 0.98 \mearth~($15\%$) uncertainty on the planetary mass $M_p$ is the quadrature sum of 0.85 \mearth~propagated from the measured radial velocity semiamplitude and 0.5 \mearth~from the stellar mass uncertainty. Further spectroscopic monitoring may reduce the former, but to fully reap the benefits of the radial velocities, the 12\% error on $M_{\star}$ should be improved. The current mass estimate is derived from an 2MASS photometry (2\% uncertainty), an empirical $M_K$-mass relation \citep[$\sim10\%$ scatter;][]{delfosse.2000.amvmsiimr} and the published system parallax \citep[$77.2 \pm 5.4$ mas;][]{van-altena.1995.gctsp}. Both improving $M_K$-mass relation for low-mass dwarfs and confirming GJ1214's parallax will be necessary to reduce GJ1214b's mass uncertainty.

The 0.12 \rearth~($5\%$) uncertainty on the planetary radius $R_p$ is already dominated by the $5\%$ uncertainty in the stellar radius $R_{\star}$, rather than the light curve parameters. This is currently constrained by the combination of the stellar density $\rho_{\star}$ that is measured directly from transit light curves \citep{seager.2003.uspspfeptlc} and the estimated mass. Even though $R_{\star} \propto (M_{\star}/\rho_{\star})^{1/3}$, and so is relatively insensitive to uncertainty in $M_{\star}$, if we fix $M_{\star} = 0.157 \pm 0.019$\msun~we find the errors on $R_{\star}$ and $R_p$ shrink by a factor of two. Improving the estimate of GJ1214's mass is also the best way to improve our measurement of the radius of the planet.

We reiterate here that our measured out-of-transit photometric modulation probes only the spatially asymmetric component of the stellar spot distribution rotating around the star. If the star hosts a subtantial, unchanging, spatially symmetric population of unocculted spots, it will bias estimates of the true planetary radius \citep{czesla.2009.saaseep} too high. If the symmetric and asymmetric components of the spots are comparable, such a bias will be at the percent level of in $R_p$, smaller than the current uncertainties.

\subsection{Metallicity of GJ1214}

Several authors have recently developed empirical photometric calibrations to estimate M dwarf metallicities from absolute ${\rm M_K}$ magnitude and $V - K$ color \citep{bonfils.2005.mdipcimrbms, johnson.2009.mrdwp, schlaufman.2010.ppcdm}. As exoplanet surveys like MEarth lavish more attention on M dwarfs as exoplanet hosts, such studies hope to address whether the giant planet vs. stellar metallicity correlation seen for AFGK stellar hosts \citep{fischer.2005.pc, johnson.2010.gposmp} extends to smaller planets and smaller stars. \citet{sousa.2008.spshpspsffe} suggest the correlation does not persist down to Neptune-mass planets, but more data are needed.

Our improved estimate of $V = 14.71 \pm 0.03$ differs significantly from the $V=15.1$ central value published in \citet{charbonneau.2009.stnls} but was the value used in the most recent analysis of this photometric metallicity calibration by \citet{schlaufman.2010.ppcdm}, who found [Fe/H] = +0.28 for GJ1214. This analysis agrees quite well with work by \citet{rojas-ayala.2010.mmphmwks} that estimates empirically calibrated metallicities from alkali metal lines in moderate resolution $K$-band spectra and finds [Fe/H] = $+0.39 \pm 0.15$ for GJ1214. This lends incremental evidence towards the persistence of the mass-metallicity correlation down to super-Earths around M dwarf hosts.

\section{Conclusions}

We have measured long-term photometric variability on GJ1214 to have a 1\% peak-to-peak amplitude { in the MEarth bandpass (715-1000 nm)} and a long rotation period, most likely an integer multiple of 53 days. Fitting very high precision light curves from the VLT, we find likely instances of GJ1214b crossing small spots during transit. Treating these occultation events as correlated noise, we find parameters for the planetary system that are consistent with previous work. 

We estimate the amplitude of time-variable changes in the apparent radius of the planet due to the observed stellar variability as $\Delta D_{\rm spots} (t) = 0.0001$ and place an upper limit of $\Delta D_{\rm spots}(\lambda) < 0.0003$ on possible spot-induced spectral features in the planet's transmission spectrum. Stellar spots do not limit current studies \citep[e.g.][]{bean.2010.temp, desert.2011.oemrasg}, but could be important for future studies of GJ1214b with {\em JWST}.

Using two years of MEarth data, we have placed limits on the presence of other transiting planets around GJ1214. With 90\% confidence, we rule out the presence of Neptune-radius transiting planets in orbits shorter than 10 days but cannot place strong constraints on planets smaller than 2.0 \rearth~at such long periods. In a system where a 1.0\mearth~planet in a 2:1 mean motion resonance would create 100 second perturbations to GJ1214b's transit times, we find no evidence for transit timing variations larger than 15 seconds. Further searches of the GJ1214 system for potentially habitable planets smaller and cooler than GJ1214b continue to be warranted.

\acknowledgments

We are extremely grateful to the KeplerCam observers Margaret Mclean, Gil Esquerdo, Mark Everett, Pete Challis, and Joel Hartman who collected V-band monitoring observations, as well as Gaspar Bakos, Dave Latham, and Matt Holman for donated time and Lars Buchhave and Xavier Bonfils for correcting the sign in the published $\gamma$ velocity. { We thank the referee for thoughtful and helpful comments that improved the paper significantly.} J.B. acknowledges funding from NASA through the Sagan Fellowship Program. The MEarth team gratefully acknowledges funding from the David and Lucile Packard Fellowship for Science and Engineering (awarded to D.C.). This material is based upon work supported by the National Science Foundation under grant number AST-0807690.  The MEarth team is greatly indebted to the staff at the Fred Lawrence Whipple Observatory for their efforts in construction and maintenance of the facility, and would like to explicitly thank Wayne Peters, Ted Groner, Karen Erdman-Myres, Grace Alegria, Rodger Harris, Bob Hutchins, Dave Martina, Dennis Jankovsky and Tom Welsh for their support.


\end{document}